\documentclass{iauc}
\usepackage{graphicx}
\pubyear{2005}
\volume{199}
\pagerange{1--}
\jname{Probing Galaxies through Quasar Absorption Lines}
\editors{P. R. Williams, C. Shu, and B. M\'{e}nard, eds.}

\title[Star formation in DLAs] 
{Star Formation and Chemical Evolution of DLAs with
Semi-Analytical Model}

\author[Hou et al.]
{J.L. Hou$^1$, C.G. Shu$^{2,1}$, S.Y. Shen$^{1}$ and R.X. Chang$^1$}

\affiliation{$^1$Shanghai Astronomical Observatory, CAS, 80 Nandan
Road,
Shanghai 200030, CHINA \\
$^2$Joint Center for Astrophysics, Shanghai Normal University,
Shanghai, 200234, CHINA \break email: hjlyx@shao.ac.cn}

\begin{document}

\maketitle

\begin{abstract}
We have examined some basic properties of damped Ly$\alpha$
systems(DLAs) by semi-analytic model. We assume that DLA hosts are
disk galaxies whose mass function is generated by Press-Schechter
formulism at redshift 3. Star formation and chemical evolution
undergo in the disc. We select modelled DLAs according to their
observational criterion by Monte Carlo simulation using random
line of sights and disk inclinations. The DLA ages are set to be 1
to 3 Gyr. By best-fitting the predicted metallicity distribution
to the observed ones, we get the effective yield for DLAs about
$0.25Z_{\odot}$. On the basis of this constrain, we further
compared our model predictions with observations at redshift 3 in
the following items: number density; gas content; HI frequency
distribution; star formation rate density; relationship between
metallicity and HI column density. We found that the predicted
number density at redshift 3 agree well with the observed value,
but the gas content $\Omega_{DLA}$ is about 3 times larger than
observed since our model predicts more DLA systems with higher
column density. The frequency distribution at higher HI column
density is quite consistent with observation while some difference
exists at lower HI end. The predicted star formation rate density
contributed by DLAs is consistent with the most recent
observations. Meanwhile, the connection between DLAs and Lyman
Break galaxies(LBGs) is discussed by comparing their UV luminosity
functions which shows that the DLAs host galaxies are much fainter
than LBGs. However, there is a discrepancy between model
prediction and observation in the correlation between metallicity
and HI column density for DLAs. Further investigations are needed
for the star formation mode at high redshift environments.

\end{abstract}

\section{Introduction}

Quasar absorption line systems are one of the best objects to
trace the physical nature of the evolving universe. Among various
absorption systems, damped Ly $\alpha$ systems (DLAs) are the
intercepting objects that have highest HI column density. It is a
common knowledge that DLAs are the progenitors of present-day
galaxies. But substantial debate continues over exactly what
populations of galaxies are responsible for them. Thanks to the
large facilities available during the past decade, more and more
data about DLAs are being provided by various observers (see
Prochaska et al. 2003).

In order to have real understanding of the physical nature of
DLAs, theorists have been using different ways to explain the
observed properties, such as simple galaxy model(Hou, Boissier \&
Prantzos 2001; Calura, Matteucci \& Vladilo 2003; Boissier,
P$\acute{e}$roux \& Pettini 2003; Lanfranchi \& Friaca 2003);
semi-analytical models(SAMs)(Cen et al. 2003; ; Cora et al. 2003;
Hou et al. 2005; Okoshi \& Nagashima 2005); and numerical
simulations (Gardner et al. 2001; Nagamine, Springel \& Hernquist
2003; Churches et al. 2004).

In this paper, we will adopt a SAM to examine in detail the
observed metallicity, HI column density and star formation
properties of DLAs in the context of standard hierarchical picture
of structure formation of the universe. The disk galaxy formation
model with single disks is adopted because we mainly concentrate
on HI column densities and the cosmic star formation rate density
contributed by DLAs rather than their kinematics. As an
illustration, DLA properties are assumed at redshift $z = 3$ and
the standard $\Lambda$CDM cosmogony is adopted.

\section{Model}

\subsection{Galaxy formation}

The galaxy formation model in this paper comes from that for disk
galaxies suggested by Mo, Mao \& White (1998, hereafter MMW). In
the model of MMW, the halo mass function at any redshift $z$ can
be described by the Press-Schechter formalism (Press \&Schschter
1974). The relation between halo mass $M$ and its circular
velocity $V_C$ is given by $ M = V_C^3/(10 G H(z))$, where  $G$ is
the gravitational constant, $H(z)$ is Hubble constant at redshift
$z$. Details about the model description can be found in MMW.

The distribution function of halo spin parameter can be described
by a log-normal function based on numerical simulation. Disks are
assumed to have exponential surface profiles $ \Sigma (R) =
\Sigma_0 \exp (-R/R_d)$, where $\Sigma_0$ and $R_d$ are the
central surface density and the scale length. The disk global
properties can be uniquely determined by parameters $\lambda$,
$V_C$, $m_d$ and the adopted cosmogony, where $\lambda$ is the
halo dimensionless spin parameter, $V_C$ is the halo circular
velocity, $m_d$ is the mass ratio of disk to halo. Here, we have
adopted the $m_d$ as a function of $V_C$ in order to consider the
galactic winds and outflow(Shu et al. 2003).

After knowing the distributions of $V_C$ and $\lambda$ for halos,
we are able to generate a sample of disk galaxies by Monte-Carlo
simulation in the $V_C$-$\lambda$ plane at redshift $z\sim 3$,
which is the parent sample for our follow-up DLA simulation.

\subsection{Star formation, chemical evolution and DLA modelling}

The adopted star formation prescription comes from disk galaxy
modelling of Boissier \& Prantzos (2000). Under the approximation
of instantaneous recycling, chemical evolution in disks can be
expressed by the simple closed-box model with metallicity $Z$ to
be $Z-Z_i = -p\,{\rm ln}\mu $, where $Z_i$ is the initial
metallicity of gas and is assumed to be $0.01Z_\odot$, $p$ is the
effective yield, and $\mu$ is the gas fraction. We assume that at
the initial time ($t = 0, z=3$), the gas surface density
$\Sigma_{g0}(R) = \Sigma_0 \exp (-R/R_d)$. Star formation proceeds
within disks in a typical time scale 1$\sim 3 \rm Gyr$ (Lanfranchi
\& Friaca 2003; Dessauges-Zavadsky et al. 2004). The effective
yield $p$ is assumed to be constant and is obtained by comparing
the metallicity distributions between model predictions and
observations for DLAs. The modelled DLAs are selected over the
sampled galaxies by random sightlines penetrating disks according
to the observed selection criterion, i.e., $N_{\rm HI} \ge
10^{20.3} cm^{-2}$. Here, random inclinations for disks in the sky
are considered.

\section{Model results and comparison with observations}

\subsection{Observations}

DLAs have shown many observational properties, including
metallicities, column densities, kinematics, etc. The observed
metallicities of DLAs adopted in the present paper mainly come
from Hou, Boissier \& Prantzos (2001) and Kulkarni \& Fall (2002)
for the Zn element. All the data presented by those authors are
compiled from the results of various observers. The observed data
of HI column densities come from the survey of Storrie-Lombardi \&
Wolfe (2000) (hereafter SW00). Since our model focuses on DLAs at
redshift $z \sim 3$, the observed properties of DLAs with redshift
$z>2 $ are selected for comparison.

\subsection{DLA host galaxies and metallicity distribution}

In Fig. 1(a), we show the resulted distribution of the impact
parameters of selected DLAs. It is found that the peak is around
3kpc, which is resulted from the huge amount of small halos in PS
formalism and the finite radius for individual galaxies that can
produce DLAs. Because the DLAs are dominated by small galaxies
which are always faint, the peak implies that the host galaxies of
DLAs are difficult to be observed photometrically.

\begin{figure}
 \includegraphics[height=5.5in,width=2.8in,angle=-90]{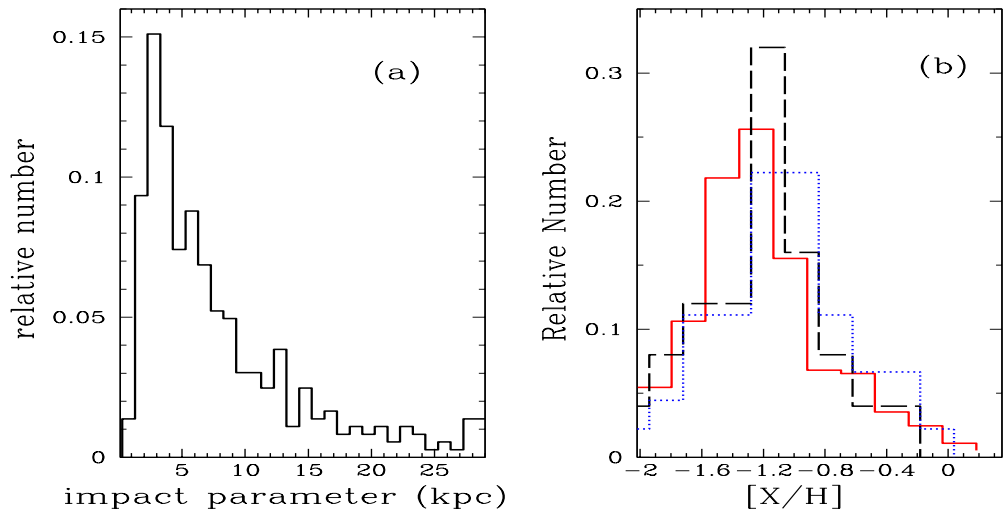}
  \caption{Distribution of impact parameter(left) and metallicity(right)
   of modelled DLA population. In the right panel, the solid, dashed and dotted
histograms denoting the model prediction, observed DLAs with $z>2$
and all DLAs, respectively.} \label{ fig:Metal}
\end{figure}

The only free parameter in the present model is the effective
stellar yield $p$, which is determined by comparing the predicted
metallicity distribution with observed one at redshift greater
than 2. The model prediction for the metallicity distributions of
DLAs is plotted as a solid histogram in Fig. 1(b) while the
observed distributions of DLAs with $z>2$ and all DLAs are plotted
as dashed and dotted histograms, respectively. We get the effect
yield $y = 0.25Z_\odot$ for the best-fit result with the
assumption of star formation timescale for DLAs being random
between 1 and 3Gyr.

\subsection{SFR density and luminosity function of DLAs}

Based on the selected DLA sample, we can get the predicted SFR
density contributed by DLAs at $z\sim 3$ and show it as a cross in
the left panel of Fig. 2 which displays the cosmic SFR density as
a function of redshift resulted from different observations. The
SFR density for DLAs with $z \gtrsim 2$ based on the $\rm CII^{*}$
absorption lines are also plotted in the figure as triangles(Wolfe
Gawsier \& Prochaska 2003, hereafter WGP03). It can be found that
the model prediction is consistent with observations and supports
the ``consensus" model described by WGP03.

\begin{figure}
 \includegraphics[height=5.5in,width=2.8in,angle=-90]{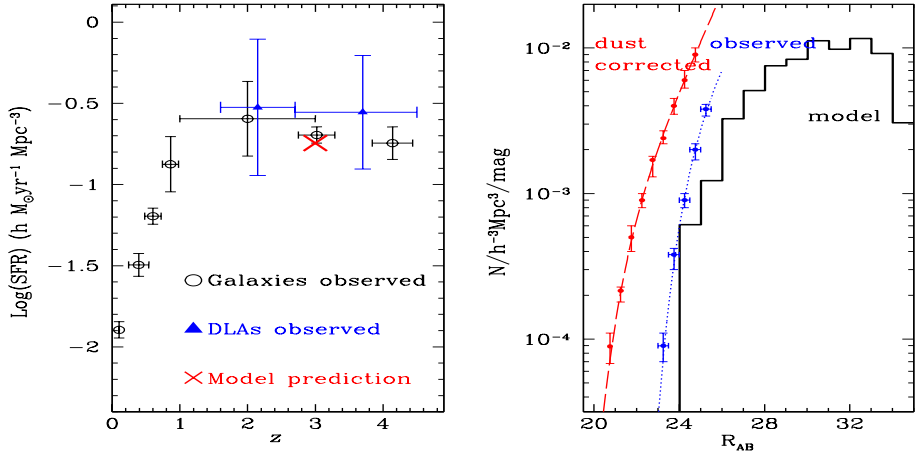}
  \caption{Cosmic SFR density as a function $z$ with the model prediction
contributed by DLAs at $z \sim 3$ as a cross (left panel). The
open circles and filled triangles with error bars are the
observational results from galaxy and from DLAs (WGP03). The right
panel shows the UV luminosity functions with the solid histogram
denoting the modelled DLAs, the dashed and dotted lines denote
LBGs with and without dust-correction which are taken from
Aldelberger \& Steidel (2000). } \label{ fig:SFR}
\end{figure}

In the right panel of Fig.2, we show, with solid histogram, the
predicted UV luminosity function of selected DLA hosts galaxies.
The observed and dust-corrected UV luminosity functions of LBGs
are also plotted as full circles (data from Aldelberger \& Steidel
2000). It can be found that the typical $R_{\rm AB}$ magnitude of
the predicted DLA hosts is $\sim 30$, which are much fainter than
LBGs with typical $R_{\rm AB}\sim 25$. This implies that a typical
DLA host galaxy in our model has its SFR about 100 times smaller
than a typical LBG. Because the number density of DLA host
galaxies is about 0.26 which is about 100 times larger than the
observed comoving number density of LBGs, our predictions of
cosmic SFR density contributed by DLAs is similar to that of LBGs
at $z \sim 3$.

\subsection{HI column density distribution and relation to metallicity}

The frequency distribution of HI column density $f(N_{\rm HI}, z)$
for DLAs , which is defined as the number of absorbers per unit
$N_{\rm HI}$ and per unit absorption distance $X$, is very
important for understanding galaxy formation and evolution in the
universe.

\begin{figure}
 \includegraphics[height=5.5in,width=2.8in,angle=-90]{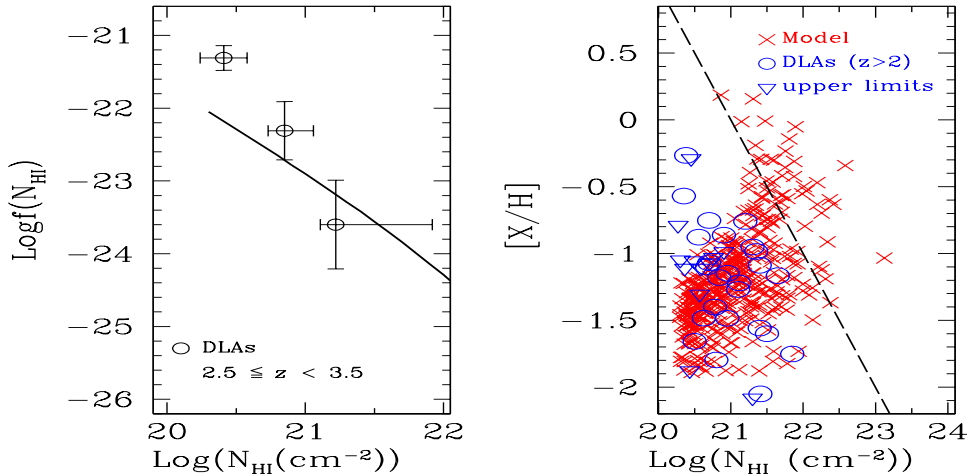}
  \caption{Left panel is $f(N_{\rm HI})$ vs $N_{\rm HI}$ for DLAs
with the solid line denoting the model prediction at $z\sim 3$,
and the circles with error bars denoting the observations for $z =
2.5\sim 3.5$ which are taken from SW00. In the right panel, we
show the predicted correlation between metallicity [X/H] and HI
column density $N_{\rm HI}$ for DLAs with crosses and open circles
denoting modelled DLAs and observations for $z>2$, respectively.
The long dashed line is ${\rm [Zn/H]+ Log}(N_{\rm HI}) = 21$.   }
\label{ fig:HI}
\end{figure}

We plot the model prediction of $f(N_{\rm HI},z)$ at $z \sim 3$ in
the left panel of Fig. 3 as a solid line while the observed
frequency distributions for DLAs with $z = 2.5 \sim$ 3.5 from SW00
is plotted as circles with error bars respectively. It can be
found from the figure that the predicted distribution agrees well
with observations at the high $N_{\rm HI}$ end. But it is smaller
than observations at low $N_{\rm HI}$ end with the maximum
difference as large as 3$\sigma$, i.e., the predicted distribution
is a bit flatter than observed ones. The similar discrepancy
appear in the more complicated numerical simulation study of
Nagamine, Springel \& Hernquist (2003) as well. This could be due
to the limitation of our simple model or perhaps this could be a
failing of the $\Lambda CDM$ power spectrum which deserves further
investigations.

Another unusual property of DLAs is that there exists an
anti-correlation between [Zn/H] and HI column densities, which is
independent of redshift as noticed by Boiss\'e et al.(1998), who
claimed that this anti-correlation is mainly due to observational
selection effects due to dust. In order to test this bias, Ellison
et al.(2001) have surveyed a sample of DLAs toward radio selected
quasars and found no significant difference in the HI distribution
to those optically selected quasars. Moreover, dust obscuration
has also been argued by Prochaska \& Wolfe (2002), who made a
detailed analysis of dust extinction contained in DLAs. They found
that inferred extinction values and apparent magnitudes imply dust
obscuration plays a relatively minor effect in the DLA analysis at
least for $z>2$.

We examine the predicted correlation between metallicity and HI
column density for the selected DLA sample in Fig.3(right panel)
with the observed results of DLAs at $z > 2$ by open circles. As
expected, model predictions show an opposite trend compared with
observations. If we apply the proposed bias of ${\rm [Zn/H]+
Log}(N_{\rm HI}) > 21$ in Fig.3 (long dashed line), to exclude the
points above this line, the difference still exists. This means
that the suggestions of Boiss\'{e} et al (1998) did not help a lot
in alleviating the discrepancy in our model.

Another possibility could be the inadequacy of the adopted Schmidt
type star formation prescription in the model. Even for nearby
galaxies, the physical basis of star formation is still poorly
known. Observers have shown various empirical prescriptions for
star formation in spirals (Kennicutt 1998; Wong \& Blitz 2002),
and most of galaxies can be fitted by a Schmidt type law. This has
been widely applied in models of galaxy evolution. In fact, gas
surface density ($\Sigma_g$) includes both the contributions of HI
($\Sigma_{\rm HI}$) and $\rm H_2$ ($\Sigma_{\rm H_2}$) with
$\Sigma_{\rm HI}$ being dominant but the correlation between these
two is different from galaxy to galaxy. Recent observations of
star formation regions in nearby galaxies done by Wong \& Blitz
(2002) showed a complex relationship between SFR and $\Sigma_{\rm
HI}$. For their spiral galaxies sample (biased to molecule-rich
galaxies), SFR shows virtually no correlation with $\Sigma_{\rm
HI}$, suggesting a maximum HI column density around $10^{21}
cm^{-2}$. This is very instructive to the star formation history
for DLAs, where the observed HI column density seems have an upper
limit.

\section{Summary}

We show in this contribution that the Semi-Analytical Model based
on the galaxy formation and evolution framework of hierarchical
structure formation scenario is quite successful in understanding
the DLAs properties. Our simple model could well reproduce most of
the observed DLA properties, such as metallicity and column
density distributions, star formation rate density contributed by
DLAs and so on. However, we found that as long as the Kennicutt
star formation prescription is adopted, model always predict a
positive correlation between HI column density and metallicity,
opposite to that observed. Different explanations exist, for
example, observational bias and dust obscuration. We suggest that
the observed trend of anti-correlation could most probably be
physical (see also Schaye 2001). A most promising explanation
might be the inadequacy of Kennicutt star formation law in the
high redshift environment. More observations and theoretical
investigations are needed to clarify this trend in the future.

\begin{acknowledgments}
We are grateful to H.J. Mo and S. Boissier for their useful
discussions and comments. This work is supported by NSFC10173017,
10133020 and 10073016, NKBRSF 1999075404 and NSC91-2112-M008-036.

\end{acknowledgments}

{}


\begin{thebibliography}{}  
\bibitem[]{} Adelberger K.L. \& Steidel C.C. 2000, \textit{ApJ}, 544, 218
\bibitem[]{} Boiss\'e P., Le Brun V., Bergeron J. \& Deharveng J.M. 1998, \textit{A\&A}, 333, 841
\bibitem[]{} Boissier S., P\'eroux C. \& Pettini M. 2003, \textit{MNRAS}, 338, 131
\bibitem[]{} Boissier S. \& Prantzos N. 2000, \textit{MNRAS}, 312, 398
\bibitem[]{} Calura F., Matteucci F. \& Vladilo G. 2003, \textit{MNRAS}, 340, 59
\bibitem[]{} Cen R., Ostriker J.P., Prochaska J.X. \& Wolfe A.M. 2003, \textit{ApJ}, 598, 741
\bibitem[]{} Churches D.K., Nelson A.H. \& Edmunds M.G. 2004, \textit{MNRAS}, 347, 1234
\bibitem[]{} Cora S.A., Tissera P.B., Lambas D.G. \& Mosconi M.B. 2003, \textit{MNRAS}, 343, 959
\bibitem[]{} Dessauges-Zavadsky M., Calura F., Prochaska J.X., D'Odorico S. \& Matteucci F.
             2004, \textit{A\&A}, 416, 79
\bibitem[]{} Ellison S., Yan L., Hook I., Pettini M., Wall J. \& Shaver P.
             2001, \textit{A\&A}, 379, 393
\bibitem[]{} Gardner J.P., Katz N., Weinberg D.H. \& Hernquist L. 2001, \textit{ApJ}, 559, 131
\bibitem[]{} Hou J.L., Boissier S. \& Prantzos N. 2001, \textit{A\&A}, 370, 23
\bibitem[]{} Hou J.L., Shu C.G., Shen S.Y., Chang R.X., Chen W.P. \& Fu C.Q. 2005, \textit{ApJ},
             624, 561 (astro-ph/0501603)
\bibitem[]{} Kennicutt R. 1998, \textit{ApJ}, 498, 541
\bibitem[]{} Kulkarni V.P. \& Fall S.M. 2002, \textit{ApJ}, 580, 732
\bibitem[]{} Lanfranchi G.A. \& Friaca A.C.S. 2003, \textit{MNRAS}, 343, 481
\bibitem[]{} Mo H.J., Mao S.D.\& White S.D.M. 1998, \textit{MNRAS}, 295, 319(MMW)
\bibitem[]{} Nagamine K., Springel V. \& Hernquist L. 2003, \textit{MNRAS}, 348, 435
\bibitem[]{} Okoshi K. \& Nagashima M. 2005, \textit{ApJ}, 623, in press
\bibitem[]{} Press W.H. \& Schechter P. 1974, \textit{ApJ}, 187, 425
\bibitem[]{} Prochaska J.X. \& Wolfe A.M. 2002, \textit{ApJ}, 566, 68
\bibitem[]{} Prochaska J.X., Gawiser E., Wolfe A.M., Castro S. \& Djorgovski S.G.
             2003, \textit{ApJL}, 595, L9
\bibitem[]{} Schaye J. 2001, \textit{ApJ}, 562, L95
\bibitem[]{} Shu C.G., Mo H.J. \& Mao S.D. 2003, astro-ph/0301035
\bibitem[]{} Storrie-Lombardi L.J. \& Wolfe A.M. 2000, \textit{ApJ}, 534, 552(SW00)
\bibitem[]{} Wolfe A.M., Gawsier E. \& Prochaska J.X. 2003, \textit{ApJ}, 593, 235(WGP03)
\bibitem[]{} Wong T. \& Blita L. 2002, \textit{ApJ}, 569, 157

\end{thebibliography}
\end{document}